\documentclass[aps,prl,twocolumn,floatfix,nofootinbib,preprintnumbers,superscriptaddress,amsmath,amssymb]{revtex4}

\usepackage{graphicx}
\usepackage{dcolumn}
\usepackage{bm}
\usepackage{hyperref}
\usepackage[mathlines]{lineno}

\begin{document}

\title{ Bayesian Evaluation of Incomplete Fission Yields}

\author{Zi-Ao Wang}
\affiliation{State Key Laboratory of Nuclear
Physics and Technology, School of Physics, Peking University,  Beijing 100871, China}

\author{Junchen Pei}
\email{peij@pku.edu.cn}
\affiliation{State Key Laboratory of Nuclear
Physics and Technology, School of Physics, Peking University,  Beijing 100871, China}

\author{Yue Liu}
\affiliation{State Key Laboratory of Nuclear
Physics and Technology, School of Physics, Peking University,  Beijing 100871, China}

\author{Yu Qiang}
\affiliation{State Key Laboratory of Nuclear
Physics and Technology, School of Physics, Peking University,  Beijing 100871, China}

\begin{abstract}
Fission product yields are key infrastructure data for nuclear applications in many aspects.
It is a challenge both experimentally and theoretically to obtain accurate and complete energy-dependent fission yields.
We apply the Bayesian neural network (BNN) approach to learn existed fission yields and predict unknowns with uncertainty quantification.
We demonstrated that BNN is particularly useful for evaluations of fission yields when incomplete experimental data are available.
The BNN results are quite satisfactory on  distribution positions and energy dependencies of fission yields.
\end{abstract}

\maketitle


\emph{Introduction.}---
Nuclear fission data is the key ingredient in many nuclear applications ~\cite{wp} such
as nuclear energy, radiation shielding, management of nuclear wastes, and production of rare isotopes.
The role of fission is also essential in synthesizing superheavy elements~\cite{she,pei}, and understandings of reactor neutrinos~\cite{neutrino} and \textit{r}-process nucleosynthesis in neutron star mergers~\cite{eichler}.
In particular, high-precision and reliable neutron-induced fission product yield (FPY) distributions of actinides are very valuable.
Systematic analysis of FPY presented
interesting insights in the evolution of nuclear structures and dynamics~\cite{itkis,andrey,scamps}.
However,
experimental measurements of FPY with continuous incident neutron energies is extremely difficult and insufficient.
In major nuclear data libraries (ENDF~\cite{endf}, JENDL~\cite{jendl}, JEFF~\cite{jeff}, CENDL~\cite{cendl}, etc.),  complete evaluations of FPY are only available for neutron incident energies
around thermal energies, 0.5 MeV and 14 MeV.  Therefore, the prediction and evaluation of incomplete FPY at other energies for fast reactors
are very anticipated.

The theoretical description of fission observables is well known as one of the most challenging tasks in nuclear physics~\cite{schunck}.
The recent developments of fully microscopic nuclear fission models such as Time-dependent Hartree-Fock-Bogoliubov ~\cite{tdhfb} and Time-dependent Generator-Coordinate Method~\cite{tdgcm,tdgcm2},
 are promising but are not ready yet for accurate quantitative applications.
There are phenomenological and semi-microscopic fission models which can well describe existing data in some region but suffers from predicting power as
fission modes evolve~\cite{gef1}. Indeed, the details of fission observables mainly depend on the multi-dimensional potential energy surfaces which are rather complex~\cite{schunck}.
 The fission of compound nuclei involve fades of quantum effects as excitation energies increase~\cite{pei,randrup}.
It is more sophisticated to describe the energy dependence of FPY ~\cite{randrup,zhao} and post-neutron FPY (independent FPY)~\cite{ifpy}.
Furthermore, the uncertainty quantification of nuclear models
has become a pressing issue in recent years~\cite{mcdonnell}.

The machine learning is a very powerful tool to learn complex big data and then make predictions.
In this respect, the Bayesian neural networks can naturally solve ill-inversed regression problems with uncertainty quantifications~\cite{witek18}.
There are a large number of experimental measurements
of fragment distributions of different nuclei and excitation energies.
To this end, the BNN is ideal for describing complex fission observables and capturing statistical properties of fission of compound nuclei.
BNN has several applications in nuclear physics for predictions of binding energies~\cite{witek18,liu,jorge,zhang}, which are useful for the inference
of unknown nuclear masses near drip lines~\cite{witek19}. It has also been used in other problems such as in simulating nuclear reaction cross sections~\cite{talou} and
equation of state~\cite{pratt}.
The Gaussian process can also solve regression problems but focuses on local correlations~\cite{witek18}.
The conventional evaluation of fission yields mainly rely on the least-squares adjustments of parameters of various phenomenological models~\cite{england}, such as
the Brosa model~\cite{brosa}. This kind of evaluations could not be applicable when very few experimental data are available.
The main objective of this work is to evaluate incomplete independent FPY based on BNN with uncertainty quantification.
The Talys ~\cite{talys} code which includes the Brosa~\cite{brosa} and GEF~\cite{gef} models has been extensively used for evaluations of fission data.
We also build BNN+Talys as an attempt to improve the descriptions of fission yields.
It is possible that the prediction and evaluation by learning of existing data can acquire underlying correlations beyond
theoretical fission models.

\emph{The models.}---
The BNN approach performs posterior inference by treating network weights as random parameters~\cite{bnn}.
With given data and finite training steps, the BNN can offer full uncertainty qualification of parameters as well as inferences through the confidence interval (CI) which
includes the true value with a probability.
The prior distribution allows
one to encode problem-specific beliefs as well as general
properties about weights, which leads to penalized functions to avoid overfitting problems.
In contrast, the frequentist inference aims to find exact parameters after infinite samplings.

We adopt a feed-forward neural network defined as,
\begin{equation}
f(x,\theta) = a + \sum_{j=1}^{H} b_j \tanh (c_j+\sum_{i=1}^{l} d_{ij}x_i)
\label{(e1)}
\end{equation}
where $H$ denotes layers of the net, $l$ denotes neurons in each layer, and the model parameters (or ``connection weights") are  $\theta$=$\{a, b_j, c_j, d_{ij} \}$.
The choice of the number of neurons defines the complexity of the networks and depends on the amount of dataset.
The inputs of the network are given by $x_i$=$\{Z_i, N_i, A_i, E_i\}$, which include the charge number $Z_i$ and neutron number $N_i$ of the fission nucleus,
the mass number $A_i$ of the fission fragment, and the excitation energy of the compound nucleus $E_i$=$e_i+S_i$ ($e_i$ and $S_i$ are neutron incident energy and neutron separation energy respectively).
The likelihood function $p(D|\theta)$ and objective function  $\chi^2(\theta)$ are given by
\begin{equation}
p(D|\theta) = \exp(-\chi^2/2),~~\chi^2 = (t_i-f(x_i,\theta)^2)/\Delta t_i^2
\end{equation}
where the data is given as $D$=$\{x_i,t_i\}$, in which $x_i$ are the inputs and $t_i$ is the output fission yield.
The posterior distribution $p(\theta|x,t)$ is obtained by
\begin{equation}
p(\theta|D)=\frac{ p(D|\theta) p(\theta)  }{  \int p(D|\theta) p(\theta) d\theta }
\end{equation}
where $p(\theta)$ is the assumed prior distribution of parameters and usually adopts a gaussian function with a width as described in Ref.\cite{bnn}.
The denominator of the integral is called ``evidence" which can be used for model comparisons.
The parameter learning is actually a maximizing posterior process via Stochastic Gradient Langevin Dynamics~\cite{bnn}.
The prediction on new inputs $x_n$ can be obtained with the net function $f(x_n,\theta)$.
The averaged prediction of BNN invokes high-dimensional intergral and can be obtained by Hybrid Markov Chain Monte Carlo integral~\cite{bnn}
over $\theta$,
\begin{equation}
<f(x_n,\theta)>=\int f(x_n,\theta)p(\theta|D)d\theta
\end{equation}
When new observed data $D_n$ are added, the posterior can be updated as
\begin{equation}
p(\theta|D,D_n)= \frac{ p(D_n|\theta)p(\theta|D)}{ \int p(D_n|\theta)p(\theta|D)d \theta}
\end{equation}
The inference can be continuously improved with the updated posterior.

\begin{figure}[t]
  \includegraphics[width=0.45\textwidth]{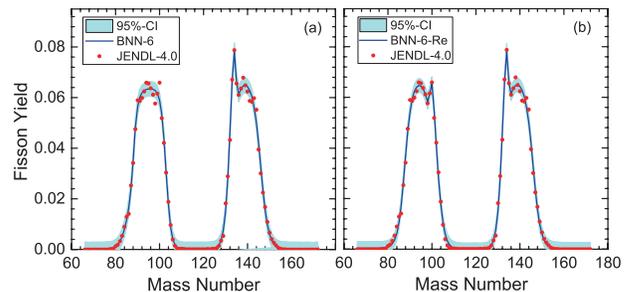}\\
  \caption{(Color online)  The BNN learning of fission yields of n+$^{235}$U at energy of 0.5 MeV from JENDL~\cite{jendl}, (a) use 6 neurons and (b) use 6 neurons and resampling.
  The shadow region corresponds to CI of BNN estimated at 95$\%$. }
  \label{fig1}
\end{figure}

\emph{Single fission}---
Firstly, to test the performance of the BNN approach, the learning of independent mass distributions of the neutron-induced fission of $n$+$^{235}$U
with energy of 0.5 MeV are studied. The  training dataset is taken from the JENDL~\cite{jendl}. In this case, the network input has only one variable $x_i$=$A_i$.
In Fig.\ref{fig1}(a), the results with 6 neurons for $N$=107 points are satisfactory with a total $\chi_N^2$ of 2.25$\times$10$^{-6}$, after 100 000 BNN samplings,
where $\chi_N^2$ is defined as $\sum\limits_i (t_i-f(x_i))^2/N$.
The largest deviations appear at mass number of 98 and 100.
It is known that global optimization and overfitting are the most challenging problems for the neural networks approach.
The BNN with a complex network can have problems in numerical convergence~\cite{witek18}.
To improve the performance of the network, we resample the points which have large deviations in the learning set and repeat the training for several cycles, in analogy to the reinforcement learning.
The results are shown in Fig.\ref{fig1}(b) with $\chi_N^2$ of 2.05$\times$10$^{-6}$.
It can be seen that the learning in Fig.\ref{fig1}(b) has been much improved compared to Fig.\ref{fig1}(a). The learning performance with resampling is slightly better than BNN with 8 neurons in which $\chi_N^2$ is 2.10$\times$10$^{-6}$. BNN needs sufficient samplings to get numerical convergence and this is very time consuming~\cite{witek18}.
We show that the reinforcement learning is helpful to efficiently obtain global optimizations.
The associated CI with a 95$\%$ probability is small and can reasonably reflect the BNN performance.

\begin{figure}[t]
  \includegraphics[width=0.45\textwidth]{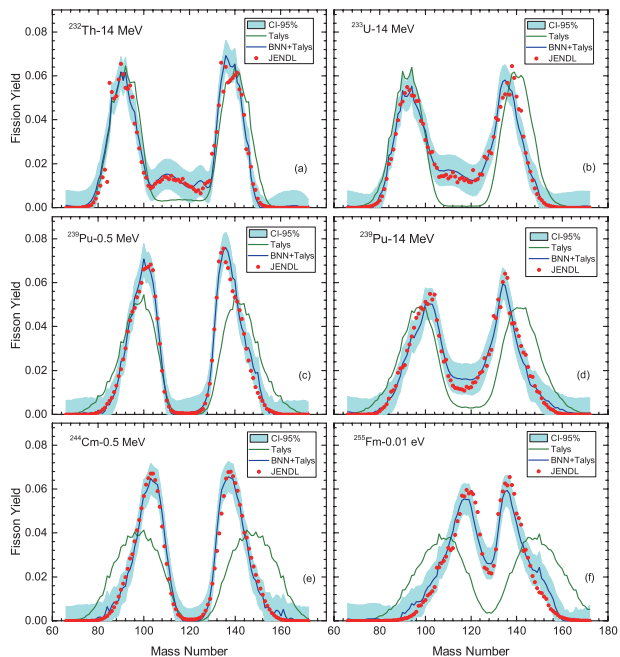}\\
  \caption{(Color online)  The BNN+Talys learning of fission yields of induced fission yields from JENDL library.   The Talys results with default parameters are also given for comparison.
  The shadow region corresponds to CI estimated at 95$\%$. }
  \label{fig2}
\end{figure}

\begin{figure}[t]
  \includegraphics[width=0.45\textwidth]{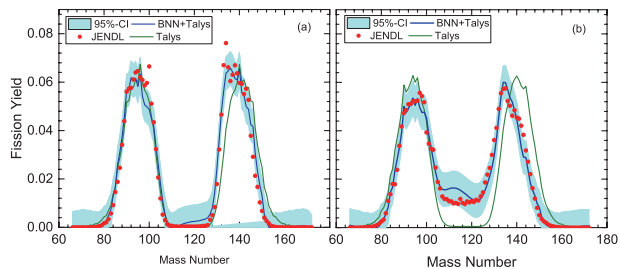}\\
  \caption{(Color online)  The BNN prediction of fission yields of n+$^{235}$U compared with JENDL, with neutron incident energy at (a) 0.5 MeV and (b) 14 MeV.
  The shadow region corresponds to the CI estimated at 95$\%$. }
  \label{fig3}
\end{figure}

\begin{table}
  \caption{ The learning and validation errors  $\chi_N^2$ ($\times10^{-5}$) of various models, in which
  BNN-32 denotes 32 neurons has been adopted in the network, and Talys(pre-n) denotes calculated pre-neutron FPY using the GEF model in the Talys-1.9 code with default parameters~\cite{talys}.}
  \label{table1}
  \centering
  \begin{ruledtabular}
  \begin{tabular}{lcc}
   Models          & learning $\chi_N^2$&  validation $\chi_N^2$   \\
\hline
BNN-32            & 1.574 & 1.317   \\
BNN-40 & 1.616 & 1.640 \\
BNN-32-resample  & 1.314 & 1.179 \\
BNN-32+Talys  & 1.902 & 1.302 \\
BNN-40+Talys & 1.139 & 1.134 \\
BNN-32-resample+Talys & 1.547 & 1.419 \\
Talys(pre-n)  & 17.56 & 8.964 \\
Talys  & 16.79 & 8.334 \\
\end{tabular}
\end{ruledtabular}
\end{table}

\begin{figure}[t]
  \includegraphics[width=0.45\textwidth]{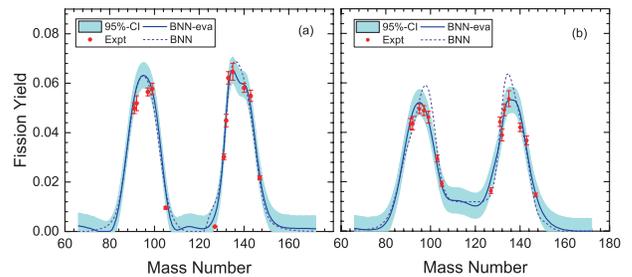}\\
  \caption{(Color online)  The BNN evaluation of fission yields of n+$^{235}$U at energies of 1.37 MeV (a) and 14.8 MeV (b), after learning
  the available experimental data~\cite{expt}.
  The dashed line denotes the BNN prediction without learning the experimental data.
  The shadow region corresponds to the CI estimated at 95$\%$. }
  \label{fig4}
\end{figure}

\begin{figure}[t]
  \includegraphics[width=0.45\textwidth]{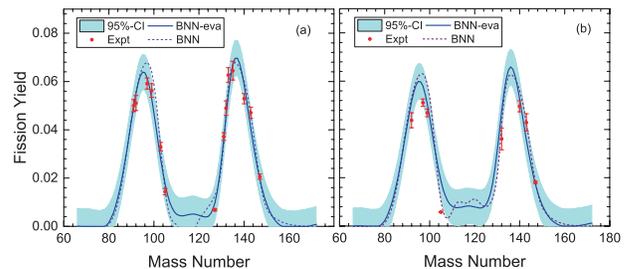}\\
  \caption{(Color online)  Similar to Fig.\ref{fig4} but for fission yields at energies of 4.49 MeV (a) and 8.9 MeV (b).}
  \label{fig5}
\end{figure}

\begin{figure}[t]
  \includegraphics[width=0.45\textwidth]{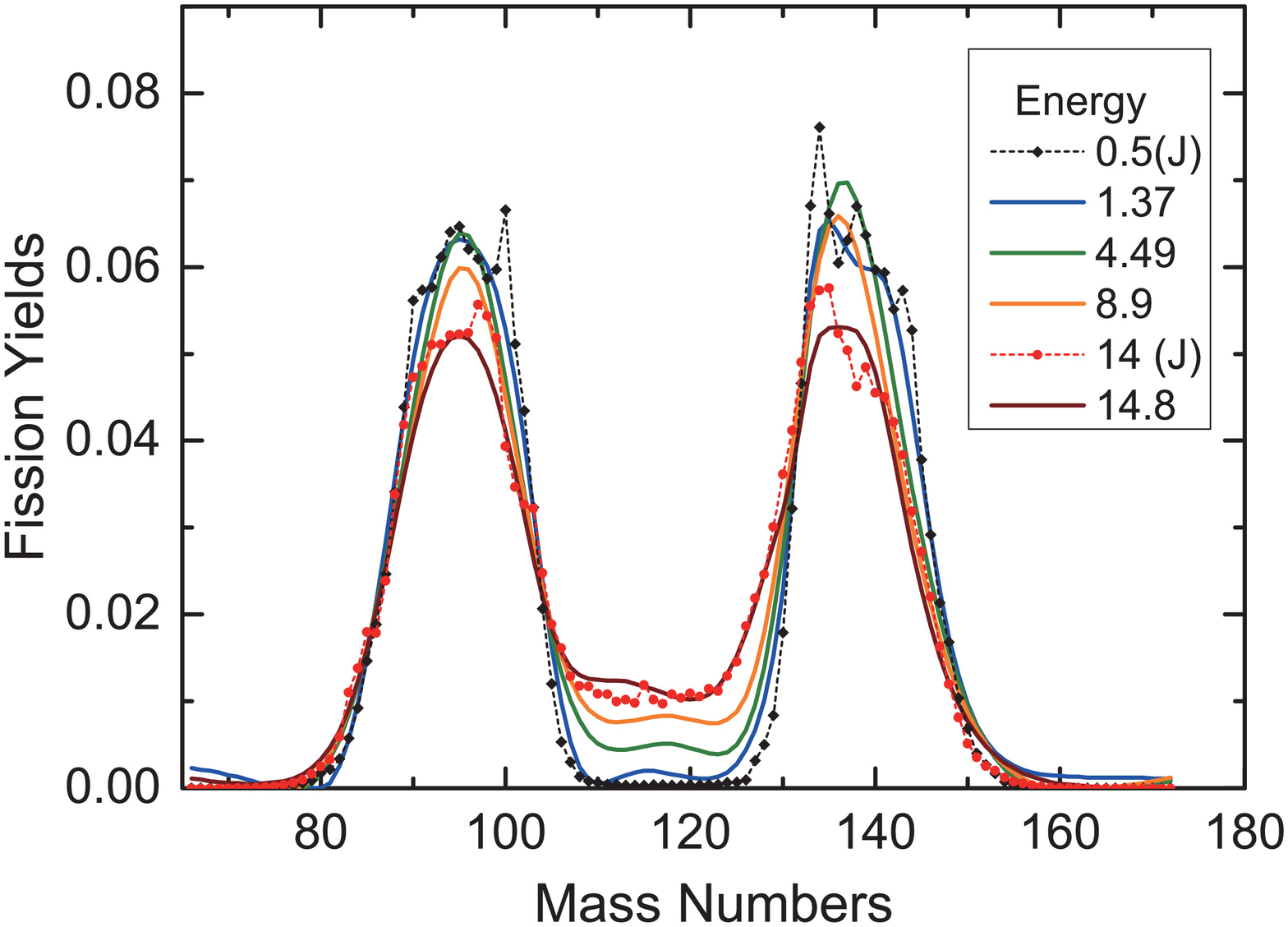}\\
  \caption{(Color online) The compiled evaluations of fission yields of n+$^{235}$U at different incident energies. The evaluations at 0.5 and 14 MeV are
  taken from JENDL~\cite{jendl}, and BNN evaluations at other energies are taken from Fig.\ref{fig4} and Fig.\ref{fig5}.}
  \label{fig6}
\end{figure}

\emph{Validation}---
We next include the evaluated experimental neutron-induced independent FPY of 30 nuclei ($^{227,229,232}$Th, $^{231}$Pa, $^{232,233,234,235,236,237,238}$U,
$^{237,238}$Np, $^{238,239,240,241,242}$Pu, $^{241,243}$Am, $^{242,243,244,245,246,248}$Cm, $^{249, 251}$Cf, $^{254}$Es, $^{255}$Fm ) from JENDL~\cite{jendl}
in the learning set. To validate the predicting power of BNN, the data of $^{235}$U has been excluded in the learning set.
To provide some physics guides, the widely used GEF model for pre-neutron FPY in the Talys-1.9 code~\cite{talys} has been adopted to obtain initial fission yields.
BNN is used to learn the residuals. The combined BNN+Talys approach is similar to
the hybrid BNN approach used for nuclear mass predictions~\cite{witek18,liu,jorge}. The network adopts 32 neurons
 for totally about 5029 data points and additional
resampled points. The dataset is much larger than that used in BNN for nuclear mass studies~\cite{witek18,liu,jorge}.
Some training results are shown in Fig.\ref{fig2}. It is shown that the Talys code with default parameters has large discrepancies compared to evaluated data.
The BNN+Tayls can remarkably reproduce the overall evaluated fission yields.
The crucial test is the validation of the BNN approach for  $n$+$^{235}$U, as shown in Fig.\ref{fig3}.
We see that the predictions of BNN+Tayls are satisfactory for  $n$+$^{235}$U at energy of 0.5 MeV and 14 MeV.
The energy dependence and the position of distributions can be well described by BNN.
This success is mainly because the learning set has included
 neighbor U and Pu isotopes.
Note that the predictions are less satisfactory around $^{227}$Th and $^{255}$Fm where neighbor nuclei in the learning set are not sufficient. The validation of $^{235}$U is not as good as the learning
in Fig.\ref{fig1}. This has also been reflected by the larger CI in Fig.\ref{fig3} compared to Fig.\ref{fig1}.

The performances of different models in learning (29 nuclei) and validation ($^{235}$U) are compared in Table \ref{table1},  with listed errors  $\chi_N^2$.
For the pure BNN approach, the best performance is BNN-32-resample with 32 neurons and resampling.
We demonstrated that the resampling is helpful in both learning and validation.
The BNN approach is not always improved with increasing numbers of neurons and local optimizations are more likely to happen.
For the BNN+Talys approach, the best performance is BNN-40. For BNN-32, the combination of Talys doesn't gain performance due to its large discrepancies.
Note that the default Talys calculations are not as good as the updated GEF model in Ref.~\cite{gef}.
In contrast, BNN plus microscopic nuclear mass models are very successful~\cite{witek18,liu,jorge}. It is expected that
BNN plus microscopic fission models~\cite{tdgcm} can further improve descriptions of fission yields.

\emph{Evaluation}---The key motivation of our BNN approach is to evaluate incomplete experimental FPY
based on the information learned from completed evaluations of other nuclei.
Fig.\ref{fig4} shows the BNN-32-resample results for fission yields of n+$^{235}$U at energies of 1.37 and 14.8 MeV after learning the JENDL library.
There are no complete evaluation in existed libraries and only a few experimental data are available~\cite{expt}.
We see that after taking into account the experimental data, the BNN can can give rather
reasonable evaluation of fission yields. The BNN predictions without learning the experimental data
are also satisfactory. This good performance is not surprise because the energies are close to 0.5 MeV and 14 MeV which are
included in the learning set.

The evaluation of n+$^{235}$U at energies of 4.49 and 8.9 MeV are shown in Fig.\ref{fig5}.
The evaluation at energies in the middle of 0.5 and 14 MeV are more challenging for BNN.
Consequently the CI in Fig.\ref{fig5} is much larger than that of Fig.\ref{fig4}.
We see that the BNN predictions have unreasonable negative FPY around mass number 110 at energies of 1.37 and 4.49 MeV.
The BNN evaluation by taking into account the experimental data can avoid the negative values.
Based on Fig.\ref{fig4} and Fig.\ref{fig5}, the fission yields from BNN at different energies are shown in Fig.\ref{fig6}.
It can be seen that around the valley (around mass number 110$\sim$120) fission yields increase monotonically as energies increase from
1.37, 4.49, 8.9, to 14.8 MeV. The two peaks corresponding to asymmetric fission modes decrease.
It is known that the symmetric fission mode will play a role as excitation energies increase~\cite{randrup,zhao}.
We demonstrated that the features of energy dependent fission yields can be successfully described by the BNN evaluation.

The BNN evaluation can reasonably give CI for uncertainty warnings.  Generally, our approach
is reliable for the distribution position but less accurate for detailed peak structures, although it can be very accurate for a single fission as shown in Fig.\ref{fig1}.
Usually fission yields are evaluated by tunning phenomenological models  which
could not be applicable when only a few experimental data are available.
In this respect, the BNN approach is superior to phenomenological evaluations.
Further improvement of our approach
is possible by adding more measured data in the learning set and adopting specialized reinforcement learning scheme.
Other auxiliary variables could be adopted for improvements. For example, the odd-even effect in charge distributions is significant~\cite{andrey}.
The physics guides on priors from microscopic fission models are also anticipated.

\emph{Summary}---
We applied BNN to learn and predict independent fission yields of actinide nuclei for the first time.
In many cases, the experimental distributions of neutron-induced fission yields are rather incomplete.
The BNN evaluation of the incomplete fission yields based on learned information is very valuable.
The BNN results are quite satisfactory regarding the distribution positions and energy dependencies of fission yields,
 while phenomenological evaluations could not be applicable when very few experimental data are available.
The BNN with resampling can improve both learning and prediction, indicating that specialized reinforcement learning is needed.
The associated confidence interval can reasonably
estimate the evaluation uncertainty. Further improvement of the BNN approach is appealing towards modeling of
reliable and quantitive nuclear fission data for practical nuclear applications.

\begin{acknowledgments}
We thank useful comments by W. Nazarewicz.
 This work was supported by  National Key R$\&$D Program of China (Contract No. 2018YFA0404403),
 and the National Natural Science Foundation of China under Grants No.11790325,11835001.
We also acknowledge that computations in this work were performed in Tianhe-1A
located in Tianjin and Tianhe-2
located in Guangzhou.
\end{acknowledgments}

\nocite{*}


\begin{thebibliography}{999}

\bibitem{wp}
L. Bernstein, D.  Brown, A. Hurst, J. Kelly, F. Kondev, E.  McCutchan, C. Nesaraja, R. Slaybaugh, A. Sonzogni, LLNL-CONF-676585,
arXiv:1511.07772.



\bibitem{she}
J.H. Hamilton, S. Hofmann, and Y.T. Oganessian, Ann. Rev. Nucl. Part. Sci. 63, 383(2013).


\bibitem{pei}
J.C. Pei, W. Nazarewicz, J.A. Sheikh and A.K. Kerman, Phys. Rev. Lett. 102, 192501(2009).

\bibitem{neutrino}
Th. A. Mueller, D. Lhuillier, M. Fallot, A. Letourneau, S. Cormon, M. Fechner, L. Giot, T. Lasserre, J. Martino, G. Mention, A. Porta, and F. Yermia,
Phys. Rev. C 83, 054615(2011).

\bibitem{eichler}
M. Eichler, et al., Astrophys. J. 808 30 (2015).

\bibitem{itkis}
M.G.Itkis, E.Vardaci, I.M.Itkis, G.N.Knyazheva, E.M.Kozulin, Nucl. Phys. A 944, 204(2015).

\bibitem{andrey}
A.N. Andreyev, K. Nishio and K.-H. Schmidt,  Rep. Prog. Phys. 81 016301(2018)

\bibitem{scamps}
G. Scamps, C. Simenel, Nature 564, 382(2018).


\bibitem{endf}
M.B. Chadwick, et al., Nuclear Data Sheets 112,  2887 (2011)

\bibitem{jendl}
K. Shibata, et al., J. Nucl. Sci. Tech. 48, 1(2011).

\bibitem{jeff}
Joint Evaluated Fission and Fusion (JEFF)
Nuclear Data Library, https://www.oecd-nea.org/dbdata/jeff/

\bibitem{cendl}
Z.G. Ge, Z. X. Zhao, H. H. Xia, Y. X. Zhuang, T. J. Liu, J. S. Zhang and H. C. Wu,
J. Korean Phys. Soc. 59, 1052 (2011)

\bibitem{schunck}
N. Schunck, L.M. Robledo, Rep. Prog. Phys. 79, 116301 (2016).

\bibitem{tdhfb}
A. Bulgac, P. Magierski, K. J. Roche, and I. Stetcu,
Phys. Rev. Lett. 116, 122504 (2016).


\bibitem{tdgcm}
D. Regnier, N. Dubray, N. Schunck, and M. Verriere,
Phys. Rev. C 93, 054611(2016)

\bibitem{tdgcm2}
W. Younes, D. M. Gogny, J.-F. Berger,
``A Microscopic Theory of Fission Dynamics Based on the Generator Coordinate Method",  Lecture Notes on Physics 950, Springer, 2019.

\bibitem{gef1}
K.-H. Schmidt and B. Jurado, Rep. Prog. Phys. 81, 106301(2018)

\bibitem{randrup}
J. Randrup and P. M\"{o}ller,
Phys. Rev. C 88, 064606 (2013).

\bibitem{zhao}
J. Zhao, T. Nik\v{s}i\'{c}, D. Vretenar, S.G.Zhou,
Phys. Rev. C 99, 014618 (2019)

\bibitem{ifpy}
S. Okumura, T. Kawano, P. Jaffke, P. Talou, S. Chiba,  J. Nucl. Sci. Tech. 55, 1009(2018).

\bibitem{mcdonnell}
J.D. McDonnell, N. Schunck, D. Higdon, J. Sarich, S.M. Wild, W. Nazarewicz,
Phys. Rev. Lett. 114, 122501 (2015)


\bibitem{witek18}
L. Neufcourt, Y. Cao, W. Nazarewicz, and F. Viens,
Phys. Rev. C 98, 034318 (2018).

\bibitem{liu}
Z. Niu and H. Liang, Phys. Lett. B 778, 48 (2018).

\bibitem{zhang}
H.F. Zhang, L.H. Wang, J.P. Yin, P.H. Chen, and H.F. Zhang, J. Phys. G 44, 045110 (2017).

\bibitem{jorge}
R. Utama, J. Piekarewicz, and H.B. Prosper, Phys. Rev. C 93, 014311 (2016);
R. Utama and J. Piekarewicz, Phys. Rev. C 96, 044308 (2017).

\bibitem{witek19}
L. Neufcourt, Y. Cao, W. Nazarewicz, E. Olsen, and F. Viens,
Phys. Rev. Lett. 122, 062502 (2019) .

\bibitem{talou}
P.Talou, P.G.Young, T.Kawano, M.Rising, M.B.Chadwick, Nuclear Data Sheets 112, 3054(2011).

\bibitem{pratt}
S. Pratt, E. Sangaline, P. Sorensen, and H. Wang,
Phys. Rev. Lett. 114, 202301 (2015)



\bibitem{england}
T.R. England, B.F. Rider, Evaluation and compilation of fission product yields, LA-UR-94-3106, Los Alamos National Laboratory (1993)


\bibitem{talys}
A.J.Koning, D.Rochman, Nuclear Data Sheets 113,2841(2012).

\bibitem{brosa}
U. Brosa, S. Grossmann, A. M\"{u}ller, Phys. Rep. 197, 167 (1990)

\bibitem{gef}
K.-H. Schmidt, B. Jurado, C. Amouroux, and C. Schmitt, Nuclear Data Sheets 131, 107(2016).

\bibitem{bnn}
R. Neal,
Bayesian Learning of Neural Network,
Springer, New York (1996).



\bibitem{expt}
M.E. Gooden, et al., Nuclear Data Sheets 131,  319(2016)





\end{thebibliography}

\end{document}